\documentclass[10pt]{article}

\usepackage{graphicx}
\usepackage{array}
\usepackage{cite}
\usepackage{boxedminipage}
\usepackage{amsmath}
\usepackage{amssymb}
\usepackage{amscd}
\usepackage{amsfonts}
\usepackage{amsthm}

\theoremstyle{plain}

\theoremstyle{definition}

\theoremstyle{remark}

\newcommand{\mycaption}[1]{\caption{{\em #1}}}
\newcommand{\be}{\begin{eqnarray}}
\newcommand{\ee}{\end{eqnarray}}

\renewcommand{\d}{\operatorname{d}}   

\newcommand{\Aut}{\operatorname{Aut}}

\newcommand{\cf}{{\em cf.}~}
\newcommand{\Cset}{\mathbb{C}}

\newcommand{\e}{{\operatorname e}}
\newcommand{\eg}{{\em e.g.}~}
\newcommand{\etc}{{\em etc}}
\newcommand{\eq}{{\,  := \, }}

\newcommand{\half}{\frac{1}{2}}

\newcommand{\Hset}{\mathbb{H}}
\newcommand{\ie}{{\em i.e.}~}

\newcommand{\Int}{\operatorname{Int}}

\newcommand{\Lie}{\operatorname{Lie}}

\newcommand{\Out}{\operatorname{Out}}

\newcommand{\QQ}{{\mathrm{q}}}

\newcommand{\qp}{\mathrm{Q}^{+}}
\newcommand{\qm}{\mathrm{Q}^{-}}

\newcommand{\SL}{\operatorname{SL}}

\newcommand{\SO}{\operatorname{SO}}

\newcommand{\Sp}{\operatorname{Sp}}
\newcommand{\Spin}{\operatorname{Spin}}

\newcommand{\SU}{\operatorname{SU}}

\newcommand{\transf}{h}

\newcommand{\Tors}{\operatorname{Tors}}
\newcommand{\Tset}{\mathbb{T}}
\newcommand{\Tw}{\text{{\tiny Twist}}}

\newcommand{\unit}{\mathbf{1}}

\newcommand{\Vol}{\operatorname{Vol}}

\newcommand{\Z}{\operatorname{Z}}
\newcommand{\Zset}{\mathbb{Z}}

\begin{document}

\begin{titlepage}
\begin{flushleft}
\hfill \\
\hfill November 18th, 2005
\end{flushleft}

\vspace*{20mm}

\begin{center}
{\bf \Large  Non-Geometric Magnetic Flux and \\
\vspace{2mm}
Crossed Modules } \\

\vspace*{8mm}

{Jussi Kalkkinen} \\

\vspace*{3mm}

{\em The Blackett Laboratory, Imperial College} \\
{\em Prince Consort Road, London SW7 2BZ, U.K.} \\

\vspace*{6mm}

\end{center}

\begin{abstract}
It is shown that the BRST operator of twisted $N=4$ Yang-Mills
theory in four dimensions is locally the same as the BRST operator
of a fully decomposed non-Abelian gerbe. Using locally defined
Yang-Mills theories we describe non-perturbative backgrounds that
carry a novel magnetic flux. Given by elements of the crossed
module $G \ltimes \Aut G$, these non-geometric fluxes can be
classified in terms of the cohomology class of the underlying
non-Abelian gerbe, and generalise the centre $\Z G$ valued
magnetic flux found by 't Hooft. These results shed light also on
the  description of non-local dynamics of the chiral five-brane in
terms of non-Abelian gerbes.

\vfill

\begin{tabbing}
{\em Keywords}  \hspace{3mm} \=  Non-Abelian Gerbes,
twisted Yang-Mills, M-Theory, Five-branes. \\
{\em E-mail}    \>  {\tt j.kalkkinen@imperial.ac.uk}
\end{tabbing}

\end{abstract}
\end{titlepage}

\tableofcontents
\pagebreak


\section{Introduction}

Among all the four-dimensional interacting quantum field  theories
the supersymmetric Yang-Mills theory is perhaps the best
understood. It enjoys beneficial symmetries that eliminate
infinities in perturbation theory on two different levels ---
first, as a gauge theory and, second, as a maximally
supersymmetric quantum field theory. In addition to this it turns
out that the theory is conformal, the beta-function vanishes, and
that it enjoys an exact non-perturbative symmetry, S-duality.

The underlying mathematical structure to gauge theory on a general
manifold $X$ is that of a principal $G$-bundle: fields on
overlapping neighbourhoods ${\cal U}_{i}$ and ${\cal U}_{j}
\subset X$ can differ by a gauge transformation $g_{ij} \in G$ on
the overlap ${\cal U}_{i} \cap {\cal U}_{j} = {\cal U}_{ij}$ in a
consistent way. Consistency here means that passing through a
third neighbourhood we get back to where we started
$g_{ij}g_{jk}g_{ki} = \unit$. The traditional way to describe a
physical system on $X$ is indeed in terms of locally, say on
${\cal U}_i$, defined differential equations. Sometimes this local
quality of differential equations in Physics can be misleading, as
some of the local fields should more properly be accommodated to
intersections ${\cal U}_{ij}$ rather than on ${\cal U}_{i}$. Yet
there is nothing in the differential equations in themselves to
give away this difference in character. This phenomenon occurred
for instance in \cite{Kalkkinen:1999uz} where the St\"uckelberg
field associated to a two-form turned out to be the connection
one-form of an only locally defined line bundle in the structure
of an underlying Abelian gerbe.

In this paper we investigate $N=4$ supersymmetric Yang-Mills
theory where the consistency condition $g_{ij}g_{jk}g_{ki} =
\unit$ has been relaxed, though in a controlled way. On a
non-Abelian gerbe we may indeed allow for such ``inconsistencies''
in the way in which the global structure of the theory is put
together from local pieces. We investigate in particular
configurations where $N=4$ supersymmetric Yang-Mills theory is
localised on double intersections ${\cal U}_{ij}$, and in a
generic local neighbourhood ${\cal U}_i$ the theory is a slightly
truncated version thereof. From the outset there is no reason to
wish to write down such configurations; this is, however, what
emerges by studying how the symmetries on the twisted Yang-Mills
theory can be embedded in the global structure of a non-Abelian
gerbe.

The underlying technical reason that allows us to make use of
non-Abelian gerbes in $N=4$ supersymmetric Yang-Mills theory is indeed
the observation that the BRST symmetry of a general non-Abelian
gerbe \cite{Kalkkinen:2005py}  is, with certain qualifications,
the same as the BRST symmetry of the  twisted $N=4$ supersymmetric
Yang-Mills theory
\cite{Yamron:1988qc,Vafa:1994tf,Labastida:1997vq}. The main novelty is that the global structure of the non-Abelian gerbe is loose enough to
include non-perturbative symmetries of the quantum theory. This makes it possible to describe new non-geometric super-Yang-Mills backgrounds
in field theory, where local fields in overlapping   neighbourhoods are related to each other by an S-duality transformation. In \cite{Hull:2004in} the non-perturbative symmetry was T-duality, hence the term ``non-geometric''.

There are at least two ways to interpret the new structure on the
overlaps ${\cal U}_{ij}$. The most straightforward is perhaps to
think of this as a twisting of the global fields on $X$ by some
local extra structure. This is the r\^ole played by the connective
structure of an  Abelian gerbe in Hitchin's generalised geometry,
for instance. The other approach is to interpret the new structure
as dynamical degrees of freedom localised in certain parts of the
space-time $X$. Perhaps a more familiar example of similar
behaviour is the fact that the presence of branes or other defects
introduces degrees of freedom on the worldvolume of these objects 
\cite{Freed:1999vc,Kalkkinen:2002tk,Kalkkinen:2004hs}.

Whichever point of view one wishes to take, the new structure will
give rise to non-geometric magnetic fluxes in terms of the
topological class of the gerbe. These fluxes are generalisations
of the magnetic flux found by 't Hooft by studying loop operators
in gauge theory \cite{'tHooft:1977hy,'tHooft:1979uj}. In a certain
sense the novel fluxes can be though of as non-Abelian surface
holonomies, analogously to as how 't Hooft's magnetic fluxes arise
from holonomies over closed loops.

\bigskip

The Paper is organised as follows: In Sec.~\ref{SYM}  we recall
aspects of super-Yang-Mills and the twisting procedure. In
Sec.~\ref{Gerbe} we quote the BRST symmetry of the non-Abelian
gerbe from \cite{Kalkkinen:2005py}, and show how it reduces to the
BRST symmetry of the Yang-Mills theory. In doing so it is
important to notice that this matching is functionally different
on local charts ${\cal U}_{i}$ and ${\cal U}_{j}$   from the
matching on double intersections ${\cal U}_{ij}$. In
Sec.~\ref{Duality} we consider a non-geometric  example where the
local description of the Yang-Mills theory on adjacent charts is
related by S-duality. In Sec.~\ref{Flux} we generalise 't Hooft's
magnetic flux to the non-geometric magnetic flux of a non-Abelian
gerbe that takes its values in the crossed module associated to
the gerbe. Finally, in Sec.~\ref{Chir} we comment on what
implications the present results have for modelling the local
dynamics of chiral five-branes in terms of non-Abelian gerbes.


\section{The $N=4$ supersymmetric Yang-Mills theory}
\label{SYM}

In this section  some of the basics of the  $N=4$ supersymmetric
Yang-Mills theory as a quantum theory are reviewed, including
electric-magnetic duality, and twists to topological theories.

The fields in the $N=4$ supersymmetric Yang-Mills theory based on
the Lie-group $G$ belong all to the adjoint representation of the
Lie-algebra $\Lie G$. Apart from the local gauge symmetry, also
the global automorphisms $\SU(4)_\mathrm{R}$ of the $N=4$ supersymmetry
algebra  act on these fields. The field content of the theory is
as follows:
\begin{itemize}
\item[{\bf --}]  Local gauge field $A$;
\item[{\bf --}]  Gaugino $\lambda$, ($\bar\lambda$) in the fundamental representation
$\mathbf{4}$ (resp.~$\bar{\mathbf{4}}$) of   $\SU(4)_\mathrm{R}$;
\item[{\bf --}]  Scalars $\Phi$ in the antisymmetric representation $\mathbf{6}$ of
$\SU(4)_\mathrm{R}$.
\end{itemize}

The coupling constant and the $\theta$-angle fit into the complex combination
\be
\tau &=& \frac{\theta}{2\pi}+ \frac{4\pi i}{g^{2}} ~.
\ee
In addition to the local gauge symmetry and the R-symmetry group
$\SU(4)_\mathrm{R}$,  the quantum theory is invariant under the S-duality
group $\SL(2,\Zset)$ (for simply laced gauge groups) generated by
\be
S =  \begin{pmatrix}
0 & 1 \\ -1 & 0
\end{pmatrix}
~\text{ and } ~ T =   \begin{pmatrix}
1 & 1 \\ 0 & 1
\end{pmatrix} ~.
\ee
The S-duality group acts on $\tau$ by fractional linear
transformation.

The $S$-transformation of the S-duality group generates
electric-magnetic duality transformations where the electric field
strength  $F = \d A + A \wedge A$ is replaced by its Hodge dual
$\tilde{F}  = \star  F$. In general there is no guarantee for the
existence of a corresponding magnetic gauge field $\tilde{A}$, and
this relationship holds indeed only using equations of motion and
in a suitably fixed gauge. This duality transformation  changes
also the electric gauge group $G$ itself to its magnetic dual
$G^{\mathrm{v}}$. The global structure of the quantum theory
should, therefore, involve both gauge groups, $G$ and
$G^{\mathrm{v}}$ \cite{Godd}. A more precise statement
demonstrated in \cite{Kapustin:2005py} is that the Wilson-'t Hooft
operators can be labelled by elements of the electric and magnetic
weight-lattices $(\Lambda_{w} \oplus \Lambda_{mw})/{\cal W}$
modulo the action of the common Weyl group ${\cal W}$.

Twisting in $N=2$ supersymmetric Yang-Mills was introduced in
\cite{Witten:1988ze}. There, twisting means identifying the
R-symmetry group $\SU(2)_{\mathrm{R}}$ with one of the factors of
the Euclidean spin-group $\Spin(4) = \SU(2)  \times \SU(2) $.
Generalisations to $N=4$ were proposed in
\cite{Yamron:1988qc,Marcus:1995mq}.  In these cases twisting
amounts to breaking the R-symmetry group $\SU(4)_{\mathrm{R}}$ to
an $\SU(2)_{\mathrm{R}}$, and then proceeding as above. One way to
distinguish twists is to determine how the fundamental
$\mathbf{4}$ decomposes into representations of the
four-dimensional spin-group. Up to interchanging left and right,
there are three  possibilities \cite{Vafa:1994tf}:
\begin{tabbing}
\hspace{10mm} \= {Chiral twist:} \hspace{20mm} \= $\mathbf{4}
\longrightarrow
(\mathbf{2},\mathbf{1})\oplus(\mathbf{2},\mathbf{1})$ \\
\> {Half
twist:} \>
$\mathbf{4} \longrightarrow (\mathbf{2},\mathbf{1})\oplus(\mathbf{1},\mathbf{1})\oplus(\mathbf{1},\mathbf{1})$ \\
\>
{Non-chiral twist:} \>  $\mathbf{4} \longrightarrow (\mathbf{2},\mathbf{2})$
\end{tabbing}
In this paper we shall concentrate on the chiral twist that leads
to the Vafa-Witten theory discussed in
\cite{Vafa:1994tf,Labastida:1997vq}. This twist has a residual
global symmetry $\SU(2)_{F}$ that interchanges the two copies of
$(\mathbf{2},\mathbf{1})$. In twisted Yang-Mills theory this
symmetry is explicitly broken  by assigning different ghost
numbers to members of the same multiplet.
\begin{table}
\be
\setlength{\extrarowheight}{3pt}
\begin{array}{|ccl|cc|}
\hline
\multicolumn{3}{|c|}{\text{Action of twisting on fields}}  & \multicolumn{2}{|c|}{\text{Ghost number}} \\
\textit{\small SYM} && \textit{\small TYM}& \textit{\small in TYM} & \textit{\small in gerbe}\\
\hline
A & \longrightarrow& A & 0 & 0 \\
\Phi & \longrightarrow &  B^{[2+]} \oplus \phi^{[0]} \oplus
C^{[0]}
\oplus \bar\phi^{[0]} &(0,2,0,-2) & (0,2,2,2) \\
\lambda & \longrightarrow &  \chi^{[2+]}  \oplus \tilde\psi^{[1]}
\oplus \eta^{[0]}  \oplus \zeta^{[0]} & (-1,1,-1,1) & (1,1,3,3) \\
\bar\lambda & \longrightarrow &  \psi^{[2+]}  \oplus
\tilde\chi^{[1]} & (1,-1) & (1,1) \\
*  & \longrightarrow & H^{[2+]} \oplus \tilde{H}^{[1]} &(  0,0) & (2,2)  \\
\hline
\end{array}
\ee
\mycaption{Field content of super-Yang-Mills (SYM) and its
decomposition in the twisted theory, Topological Yang-Mills (TYM);
Ghost numbers in TYM and the gerbe. Square brackets refer to the
degree a differential form, and the superscript $[2+]$ to a
self-dual two-form\label{Gh}.}
\end{table}

Re-identification of the spin-group in the quantum theory changes
the energy-momentum tensor and therefore, potentially, the
underlying quantum theory itself. In a flat or hyper-K\"ahler
background metric the twisted and the physical $N=4$
super-Yang-Mills theories are nevertheless equivalent
\cite{Witten:1994ev,Vafa:1994tf}. Examples of such four manifolds
are the compact K3, and the noncompact hyper-K\"ahler resolutions
of orbifolds of the form $M=\Cset^{2}/\Gamma$, where $\Gamma
\subset \SU(2)$ is a discrete subgroup and $\Cset^{2}$ its linear
representation.


\section{Locally twisted Yang-Mills on a gerbe}
\label{Gerbe}

The topology of a non-Abelian gerbe can be  given in terms of the
cocycle data $(g_{ijk}, \lambda_{ij})$. Here $\lambda_{ij}$ is  an
$\Aut G$ valued function on the double intersection of local
charts ${\cal U}_{ij}$, and $g_{ijk}$ is a $G$-valued function on
the triple intersection of local charts ${\cal U}_{ijk}$. The
distinction between $G$ and $\Aut G$-valued objects is quite
important. (This structure will be discussed in more detail in
Sec.~\ref{CM}.) In general an automorphism can involve an outer
part that cannot be effected by conjugation with a group element.
In the case of a Lie-group, such outer automorphisms are
symmetries of the Dynkin diagram. For instance, $\Out \SU(n) =
\Zset_{2}$ (complex conjugation, $n>2$) and $\Out \Spin(8) =
\Zset_{3}$ (triality).

A fully decomposed gerbe was described in terms of the above
cocycle  in \cite{BreenAst}, and in terms of differential geometry
in \cite{BM2}. When the decomposition is only partial, one is lead
to intermediate structures that involve local non-Abelian bundles
but whose characteristic classes are Abelian in the sense of
\cite{BreenAst}. This is true of bundle gerbes \cite{M}.

The BRST operator that generates infinitesimal symmetries of a
gerbe was  constructed in \cite{Kalkkinen:2005py}. The local
fields involve the local connection which is a $\Lie \Aut G$
valued one-form $m_{i}$; a $\Lie G$ valued one-form $\gamma_{ij}$
on the double intersection of local charts ${\cal U}_{ij}$; and a
local  $\Lie G$ valued two-form $B_{i}$. In
Refs.~\cite{BM2,Kalkkinen:2005py} these  fields are really {\em
group} valued differential forms \cite{BM1}, though for the
present discussion they reduce to algebra valued forms. To write
the BRST operator down, we need the following notation:
\begin{itemize}
\item[{\bf --}]
The adjoint action of the group element is denoted  by
$\iota_{g}(h) = g h g^{-1}$.
\item[{\bf --}]  Given an Hodge star $\star$, we denote $\iota_{x}^{+} =  {\scriptstyle \half} (1+\star) \iota_{x}$ for any Lie-algebra valued two-form $x$.
\item[{\bf --}]
The covariant exterior derivative of Lie-algebra  valued forms
is
\be
\d_{m_{i}}x &\eq& \d x + [m_{i}, x]~.
\ee
\item[{\bf --}]
The action of an automorphism $\lambda$ on an automorphism  valued
form $m$ is denoted ${}^{\lambda}m$. For connection one-form we
write ${}^{\lambda*}m$.
\item[{\bf --}]
The local field strength of $m_{i}$ is
\be
\kappa(m_{i}) &=& \d m_{i} + \half [m_{i}, m_{i}] ~.
\ee
\end{itemize}

The fields appearing in a fully decomposed  non-Abelian \cite{BM2}
gerbe and the associated BRST operator $\QQ$ are summarised in
Table \ref{fieldsTable2}.
\setlength{\extrarowheight}{3pt}
\begin{table}
\be
\begin{array}{|c|c|cccc|}
    \hline
    & \text{ghost\#}  & \text{0-form} & \text{1-form} &
\text{2-form} & \text{3-form}   \\
    \hline
    G & 0  & g_{ijk} & \gamma_{ij}  & B_i, ~\delta_{ij} & \omega_{i} \\
      & 1  & a_{ij}  & E_i, ~\eta_{ij}&  \alpha_{i}  & \\
      & 2  & \phi_{i}, ~b_{ij} & \rho_i && \\
      & 3  & \sigma_{i} &&& \\
    \hline \Aut(G)
      & 0  & \lambda_{ij} & m_i & \nu_{i}& \\
      & 1  &  c_i    &  \pi_i  && \\
      & 2  &  \varphi_{i}  &&&  \\
  \hline
\end{array}
\nonumber
\ee
\mycaption{Fields and field strengths on the universal gerbe.
\label{fieldsTable2}}
\end{table}
The BRST operator of a fully decomposed gerbe \cite{Kalkkinen:2005py} is
\be
\QQ m_{i} &=& \pi_{i} + \iota_{E_{i}} - \d_{m_{i}}c_{i}
\\
\QQ_{c} \gamma_{ij} &=& \eta_{ij}+ E_{i} - \lambda_{ij}(E_{j})
+ \d_{m_{i}}a_{ij} - [\gamma_{ij},a_{ij}]   \\
\QQ_{c} B_{i} &=& \alpha_{i} + \d_{m_{i}}E_{i}   \\
\QQ_{c} \pi_{i} &=& \iota_{\rho_{i}} + \d_{m_{i}}\varphi_{i}
  \\
\QQ_{c} E_{i} &=& -\rho_{i} + \d_{m_{i}} \phi_{i}   \\
\QQ c_{i} &=& \varphi_{i} + \iota_{\phi_{i}} + {\scriptstyle \half} [c_{i},c_{i}]
  \\
\QQ_{c} \eta_{ij} &=& - \d_{m_{i}}b_{ij} + \rho_{i} -
\lambda_{ij}(\rho_{j}) + [\iota_{\eta_{ij}} -\pi_{i} , a_{ij}]
- [\varphi_{i} + \iota_{b_{ij}}, \gamma_{ij}] \\
\QQ_{c} \alpha_{i} &=&   \d_{m_{i}}\rho_{i} - [\nu_{i},\phi_{i}] - [\pi_{i},E_{i}] - [\varphi_{i},B_{i}]   \\
\QQ_{c} \varphi_{i} &=& -\iota_{\sigma_{i}}   \\
\QQ_{c} \phi_{i} &=& \sigma_{i}   \\
\QQ_{c} \rho_{i} &=&  \d_{m_{i}}\sigma_{i} + [\pi_{i},\phi_{i}] + [\varphi_{i},E_{i}]   \\
\QQ_{c} \sigma_{i} &=& - [\varphi_{i},\phi_{i}]   \\
\QQ_{c} a_{ij} &=& b_{ij}- \phi_{i} + \lambda_{ij}(\phi_{j}) +
{\scriptstyle \half}[a_{ij},a_{ij}]   \\
\QQ_{c} b_{ij} &=&   \sigma_{i} - \lambda_{ij}(\sigma_{j})
-[\varphi_{i} + \iota_{b_{ij}},a_{ij}] ~.
\ee
Here $\QQ_{c} x $ for any field $x$ is defined as $\QQ x + [c_i,
x]$.

This BRST algebra closes on-shell \cite{Kalkkinen:2005py}. By
on-shell we mean that on double intersections ${\cal U}_{ij}$ the
relationships
\be
{}^{\lambda_{ij}*}m_{j} - m_{i} + \iota_{\gamma_{ij}} &=& 0 \\
{}^{\lambda_{ij}}c_{j} - c_{i} - \iota_{a_{ij}} &=& 0 \\
{}^{\lambda_{ij}}\pi_{j} - \pi_{i} + \iota_{\eta_{ij}} &=& 0 \\
{}^{\lambda_{ij}}\varphi_{j} - \varphi_{i} - \iota_{b_{ij}} &=& 0
~,
\ee
are imposed. There are similar relationships on triple
intersections, for a full discussion see \cite{Kalkkinen:2005py}.
Also, on-shell the BRST operator squares to the gauge
transformation
\be
\QQ^{2} x_i &=& [\varphi_i +\iota_{\phi_{i}}, x_{i}]
\ee
on any field $x_{i}$, except on $\eta_{ij}$ and $b_{ij}$.

\subsection{Isolated local charts}
\label{G}

Let us set, temporarily, the intersection fields to trivial values
\be
\gamma_{ij} = \eta_{ij} = b_{ij} = a_{ij} = 0
~,
\ee
and work of a single chart ${\cal U}_{i}$. We can therefore omit
the indices $i$ from the formulae. If this is the case, the sum of the BRST transformations $\qp$ and  $\qm$ of
Ref.~[\citen{Labastida:1997vq}, {Eq.~(2.24)}] coincide with
$\QQ$ with the following identifications:
\be
A^{{\Tw}} & = & m \\
B^{{\Tw}}  &=& {\scriptstyle  2} ~ \iota_{B}^{+} \\
C^{{\Tw}} &=& 0 \\
\psi^{{\Tw}} &=& - {\scriptstyle  \half} \pi \\
\tilde\psi^{{\Tw}} &=& {\scriptstyle \sqrt{2}} ~ \iota_{\alpha}^{+} \\
\chi^{{\Tw}} &=& - {\scriptstyle \sqrt{2} }~ \iota_{\d_{m}E}^{+} \\
\tilde\chi^{{\Tw}} &=& - {\scriptstyle \half }\iota_{E} \\
\phi^{{\Tw}} &=& {\scriptstyle \frac{1}{2\sqrt{2}}} ~ \varphi \\
\bar\phi^{{\Tw}} &=& -{\scriptstyle \frac{1}{2\sqrt{2}}} ~ \iota_{\phi} \\
\zeta^{{\Tw}} = \eta^{{\Tw}} &=& -{\scriptstyle \frac{1}{4}} ~ \iota_{\sigma} \\
\tilde{H}^{\prime{\Tw}} &=& {\scriptstyle \half} ~ \iota_{\rho} \\
H^{\prime{\Tw}} &=& - {\scriptstyle \sqrt{2}} ~ \iota_{\big(-
\d_{m} \rho + [\nu, \phi] + [\pi,E] \big)}^{+} ~.
\ee
The BRST operator splits $\QQ = \qp + \qm$, and we have (for $c_i
= 0$)
\be
\begin{array}{clccl}
\qp m =& \pi & \qquad & \qm m =& \iota_{E} \\
\qp B =& \alpha  & \qquad & \qm B =& \d_{m}E \\
\qp \pi =& \d_{m}\varphi & \qquad & \qm \pi =& \iota_{\rho} \\
\qp E =& -\rho & \qquad & \qm E =&  \d_{m}\phi \\
\qp \alpha =& - [\varphi,B] & \qquad & \qm \alpha=& \d_{m} \rho -[\nu, \phi] - [\pi,E] \\
\qp \varphi =& 0 & \qquad & \qm \varphi =& -\iota_{\sigma} \\
\qp \phi =&  \sigma & \qquad & \qm \phi =& 0  \\
\qp \rho =& [\varphi,E] & \qquad & \qm \rho =& \d_{m}{\sigma} + [\pi,\phi ] \\
\qp \sigma =& 0 & \qquad & \qm \sigma =& -[\varphi,\phi] ~.
\end{array}
\ee

The BRST operator of the non-Abelian gerbe involves also the
anti-self-dual part of the two-forms appearing above. This means
that the BRST algebra, with these identifications, forms a
self-consistent extension of the twisted algebra where only the
self-dual part appears. Note, however, that for this comparison we
had to set $C$ to zero and $\eta=\zeta$ in the twisted theory. 

The above restrictions mean that the sector we are interested in
is not quite a balanced topological quantum field theory
\cite{Dijkgraaf:1996tz}, because ghost number grading is different
(\cf Table \ref{Gh}) and we have replaced two fields $\eta$, $\zeta$ that in the twisted theory have opposite ghost number with a single field
$\sigma$. Therefore, the usual arguments for the absence of ghost
number anomaly are not quite valid. Despite our removing these
restrictions in Sec.~\ref{L}, this departure from balanced TQFT
will become even more pronounced, as certain non-local effects
will have to be  incorporated in the formalism.

\subsection{Intersections of local charts}
\label{L}

The gauge field $m$ in the non-Abelian gerbe is  not the only
one-form at our disposal, but we have also $m_{i} -
\iota_{\gamma_{ij}}$. (This is of course the same as
${}^{\lambda_{ij}*}m_{j}$.) For this to make sense we must work on
a double intersection ${\cal U}_{ij}$, and turn on all other
fields supported on double intersections as well, $\eta_{ij}$,
$a_{ij}$, and $b_{ij}$.

The BRST algebra of the gerbe turns out to be too large as such,
however, and we have to restrict $a_{ij}=0$. As then also $\QQ
a_{ij} =0$, we have the conditions
\be
a_{ij} &=& 0 \\
b_{ij} &=& \phi_i - \lambda_{ij}(\phi_j) ~.
\ee
This is the same restriction as what was necessary in
\cite{Kalkkinen:2005py} to map the nilpotent BRST  operator on the
universal gerbe to the non-nilpotent operator that implemented the
infinitesimal symmetries of a non-Abelian gerbe of \cite{BM2}. It
was shown in \cite{Kalkkinen:2005py} in particular that on-shell
these two equations can be imposed as algebraic identities.

These identities lead now to two simplifications in the
constraints:
\be
{}^{\lambda_{ij}}c_j - c_i &=& 0 \label{cc} \\
{}^{\lambda_{ij}}(\varphi_j + \iota_{\phi_j}) - (\varphi_i  +
\iota_{\phi_i}) &=& 0 ~.  \label{cf}
\ee
Then $\varphi_i + \iota_{\phi_i}$ is globally well-defined section of a vector bundle. Also,
\be
\QQ^{2}x = [\varphi + \iota_{\phi}, x]
\ee
for any field $x=B_i, \eta_{ij}$  \etc.

With these restrictions on the gerbe, the BRST operator in
Ref.~[\citen{Labastida:1997vq}, {Eq.~(2.24)}]  reduces precisely
to the BRST operator of the fully decomposed gerbe. The precise
identifications, that essentially generalise the above-presented,
are as follows:
\be
A^{{\Tw}} & = & m_i - \iota_{\gamma_{ij}} \\
B^{{\Tw}}  &=& {\scriptstyle  2} ~ \iota_{B_i}^{+} \\
\psi^{{\Tw}} &=& - {\scriptstyle  \half} (\pi_i -\iota_{\eta_{ij}}) \\
\tilde\psi^{{\Tw}} &=& {\scriptstyle \sqrt{2}} ~ \iota_{\alpha_i}^{+} \\
\chi^{{\Tw}} &=& - {\scriptstyle \sqrt{2} }~ \iota_{\d_{m_i}E_i}^{+} \\
\tilde\chi^{{\Tw}} &=& - {\scriptstyle \half }\iota_{\lambda_{ij}(E_j)} \\
\phi^{{\Tw}} &=& {\scriptstyle \frac{1}{2\sqrt{2}}} ~ \varphi_i \\
\bar\phi^{{\Tw}} &=& -{\scriptstyle \frac{1}{2\sqrt{2}}} ~ \lambda_{ij}({\phi_j}) \\
C^{{\Tw}} &=& {\scriptstyle \frac{1}{4\sqrt{2}}} ~ (\lambda_{ij}({\phi_j}) - \phi_i) \\
\zeta^{{\Tw}} &=& -{\scriptstyle \frac{1}{4}} ~ \iota_{\sigma_i} \\
\eta^{{\Tw}} &=& -{\scriptstyle \frac{1}{4}} ~ \iota_{\lambda_{ij}(\sigma_j)} \\
\tilde{H}^{\prime{\Tw}} &=& {\scriptstyle \half} ~ \iota_{\lambda_{ij}(\rho_j)} \\
H^{\prime{\Tw}} &=& - {\scriptstyle \sqrt{2}} ~
\iota_{\big(-\d_{m_i} \rho_i + [\kappa(m_i), \phi_i] - [B_i,
\lambda_{ij}(\phi_i)] + [\pi_i,E_i]  \big)}^{+} ~.
\ee

It does not seem to be possible to define the operators $\qp$ and
$\qm$ separately, as this would require making sense for instance of
\be
\qp (\phi_i - \lambda_{ij}(\phi_j)) & \stackrel{?}{=} & \sigma_i \\
\qm (\phi_i - \lambda_{ij}(\phi_j)) & \stackrel{?}{=} & -
\lambda_{ij}(\sigma_j) ~.
\ee
Only the sum is $\qp +\qm = \QQ$ is well-defined, and the topological quantum field theory is not balanced.

\subsection{Global structure}

On isolated local neighbourhoods ${\cal U}_{i}$ we  have
replicated in Sec.~\ref{G} the structure of a standard twisted
Yang-Mills theory. The minor differences that remain were
\begin{itemize}
\item[{\bf --}]  Some twisted Yang-Mills fields are constrained
\be
C &=& 0 \\
\eta &=& \zeta
\ee
so that the global flavour symmetry $\SU(2)_{F}$ is broken;
\item[{\bf --}]  The non-Abelian gerbe keeps track also of anti-self-dual
components; and
\item[{\bf --}]  The ghost number grading is compatible with $\SU(2)_{F}$
in the gerbe but not in Yang-Mills.
\end{itemize}
These restrictions are enough to break the balanced structure of
the standard twisted theory, though. On intersections of these neighbourhoods  ${\cal U}_{ij}$ the
topological theory on the gerbe is even further away from being
balanced, as the BRST operator does not split any more  $\QQ \neq
\qp +\qm$. This means that though the theory might be locally
nearly holomorphic on ${\cal U}_{i}$, its global structure is
certainly put together by using non-holomorphic rules on double
intersections ${\cal U}_{ij}$.

In the construction of Sec.~\ref{L} there was no restriction
on the fields at all, in fact all three scalars were active
\be
 {2\sqrt{2}} ~ \phi^{{\Tw}} &=&  \varphi_i \\
 {2\sqrt{2}} ~ \bar\phi^{{\Tw}} &=& - \iota_{\lambda_{ij}({\phi_j})} \\
 {4\sqrt{2}} ~ C^{{\Tw}} &=&  \iota_{\lambda_{ij}({\phi_j}) - \phi_i} ~.
\ee
Similarly, their superpartners were unconstrained
\be
-4 ~ \zeta^{{\Tw}} &=&   \iota_{\sigma_i} \\
-4 ~ \eta^{{\Tw}} &=&   \iota_{\lambda_{ij}(\sigma_j)} ~.
\ee
This construction reduces to the earlier  construction, of
course, when $\varphi_i$, $\iota_{\phi_i}$, and $\sigma_i$ are
separately covariant. This does not follow from the covariance of
$c_i$ and $\varphi_i + \iota_{\phi_i}$ observed in (\ref{cc}) --
(\ref{cf}) alone. The flavour symmetry $\SU(2)_{F}$ is broken in
this case not by ghost number assignments but rather by
\v{C}ech-degree and the local  structure of the
$(\varphi_i,\phi_i, \sigma_i)$ system.

In a local quantum field theory one would  usually expect to find
one degree of freedom per Planck volume.  In the present theory,
however, where two local constructions overlap we seem to have an
increase in the number degrees of freedom, in terms of the new
fields  $C^{\Tw}$ and $\zeta^{\Tw} - \eta^{\Tw}$. This does not
need to change the structure of the Hilbert space radically,
because the theory is after all a topological quantum theory whose
Hilbert space is expected to be finite dimensional. One can think of this data   either as a locally defined twist of
the global configuration, or as new degrees of freedom. In the
former case this data is kept fixed in the path integral, and
characterise the global configuration. In the latter case these
fields describe new degrees of freedom on the overlaps, and should
be integrated over in a path integral. All of the
new degrees of freedom on the overlaps ${\cal U}_{ij}$ with their
superpartners are summarised fully in Table \ref{Doverlap}.
\begin{table}
\be
\begin{array}{|c|c|}
\hline \text{Field} & \text{Superpartner} \\
\hline
\gamma_{ij} &\eta_{ij} \\
\lambda_{ij}(\phi_j) - \phi_i &  \lambda_{ij}(\sigma_j) - \sigma_i
\\
\lambda_{ij}(E_j) - E_i &  \lambda_{ij}(\rho_j) - \rho_i \\
\delta_{ij} &  \lambda_{ij}(\alpha_j) - \alpha_i   \\
\hline
\end{array}
\ee
\mycaption{Degrees of freedom and their superpartners on ${\cal
U}_{ij}$. \label{Doverlap}}
\end{table}

If we wish indeed to interpret these discontinuities in the various fields in Table \ref{Doverlap} as new degrees of freedom and integrate over them in a path integral, giving fields on an open cover ${\cal U}_{i}$, ${\cal U}_{ij}$, ${\cal U}_{ijk}$, and so on is clearly not
the right way to organise this data. This is because at a single point in \eg  a double overlap we have simultaneously three different sets of fields --- those defined on ${\cal U}_{i}|_{j}$,  ${\cal U}_{j}|_{i}$, and ${\cal U}_{ij}$.

The additional structure that we need in order to understand the local distribution of degrees of freedom is in fact a compatible triangulation on $X$, where every simplex $v$ of maximal dimension carries an index $i$ corresponding to a local chart  where it is included $v \subset {\cal U}_{i}$, each codimension one simplex $s$ carries similarly an index $ij$ corresponding to an overlap $s \subset{\cal U}_{ij}$, and so forth. A similar procedure leads to Gawedzki's topology on the loop space of X, and can be used to write down an explicit formula for the holonomy of an Abelian $n$-gerbe in \cite{Brylinski}. Consequently, though a field may be defined over all ${\cal U}_{ij}$, it might be physical only on codimension one simplexes $s \subset{\cal U}_{ij}$ included in the triangulation we have chosen. 
  
In this sense overlaps ${\cal U}_{ij}$ can be thought of as virtual domainwall defects in the ambient spacetime $X$. Of course, overlaps are open subsets of $X$ and a domainwall defect is usually a closed submanifold embedded in $X$, so that the two structures are quite different. The point is that were there a domainwall embedded in $X$, the degrees of freedom on it should be labelled in terms of data defined on ${\cal U}_{ij}$. To develop these ideas fully, one should find out in what extent an eventual path integral formulation of the theory really depends on such a triangulation, and whether degrees of freedom on the above codimension one simplexes really imply the presence of a physical domainwall.

It is interesting to note nevertheless that at least the Bosonic new degrees of freedom on such a virtual domainwall seem to include  degrees of freedom localised on a physical domainwall in four dimensions: a vector $\gamma_{ij}$ and a scalar $\lambda_{ij}(\phi_j) - \phi_i$. The fields here are Bosonic components of a supermultiplets on a superspace where the BRST symmetry acts by odd translations. Untwisting these supermultiplets (with the other Fermionic data on the overlap) would unfortunately seem to require more detailed knowledge  about the physical phase space, equations of motion and gauge fixing in particular.\footnote{Note that $\delta_{ij}$ should be seen as a part of the curvature of the global configuration, and that $\lambda_{ij}(E_j) - E_i$ can be absorbed in $\eta_{ij}$. Though the interpretation of these
fields must be left open at this stage, their presence  on the
overlap may reflect the intricate structure of a
non-Abelian gerbe rather than new degrees of freedom.}

The next question is the number of degrees of freedom on triple
intersections ${\cal U}_{ijk}$. We have not introduced new fields
explicitly on these overlaps, and the only object carrying three
\v{C}ech indices is the class of the gerbe $g_{ijk}$ which we keep
fixed.

In the Abelian case it is easy to check whether fields defined on
a double overlap, say $x_{ij}^\mathrm{A}$, can be accounted for
{\em locally} in terms of differences $x_{j}^\mathrm{A} -
x_{i}^\mathrm{A}$: the check is simply that $x_{ij}^\mathrm{A}$
should be closed under the  \v{C}ech coboundary operator
\be
(\partial x^\mathrm{A})_{ijk} &=& x_{ij}^\mathrm{A} +
x_{jk}^\mathrm{A} + x_{ki}^\mathrm{A} ~. 
\ee
In the non-Abelian case the situation is not quite so clear:
fields on different charts cannot be compared directly, as they
must be mapped first in the right frame using the transition
functions $\lambda_{ij}$. Given this structure one can
nevertheless define the covariant  \v{C}ech coboundary operator
\be
(\partial_{\lambda} x)_{ijk} &=& x_{ij} + {}^{\lambda_{ij}} x_{jk}
+ {}^{\lambda_{ij}\lambda_{jk}} x_{ki}   \label{parla}
\ee
and use it to check what happens to a field that is clearly a
difference of local fields, say $x_{ij}=  {}^{\lambda_{ij}}x_j  -
x_i$. Suppose $x_i$ is a differential form of positive rank. Then
the result is its commutator with the class of the gerbe
\be
(\partial_{\lambda} x)_{ijk} &=& [g_{ijk}, x_{i}] ~.
\ee
This is the consistent result, and indicates that there are no new
degrees of freedom localised on ${\cal U}_{ijk}$. To spell this
out more directly, note that changing charts over a fixed point in
${\cal U}_{ij}$ the differential form $x_i$ as expressed in terms
of $x_j$ gets shifted
\be
x_i &=&  {}^{\lambda_{ij}}x_j - x_{ij} ~.
\ee
Repeating this procedure three times  through ${\cal U}_{ij}
\longrightarrow {\cal U}_{jk} \longrightarrow {\cal U}_{ki}$ we
get
\be
x_i &=& {}^{\lambda_{ij} \lambda_{jk} \lambda_{ki}}x_i -
(\partial_{\lambda} x)_{ijk} \\
&=& x_i ~.
\ee
There is therefore no inconsistency in how $x_{ij}$, $x_{jk}$, and
$x_{ki}$ are defined, and no new degrees of freedom on ${\cal
U}_{ijk}$.

These identifications put then the class $g_{ijk}$ directly in
evidence. On a triple intersection ${\cal U}_{ijk}$ we have three
different twisted scalar fields, including $C^{{\Tw}}$. It is easy to see that the
departure of the simplifications of the local construction on
${\cal U}_{i}$ gives rise to
\be
{4\sqrt{2}} \Big(\partial_{\lambda} C^{\Tw}\Big)_{ijk} &=&
\iota_{[g_{ijk}, \phi_{i}]} \\ 4 \Big(\partial_{\lambda}
(\zeta^{\Tw} - \eta^{\Tw}) \Big)_{ijk} &=& \iota_{[g_{ijk},
\sigma_{i}]} ~.
\ee
The same calculation for all new fields on double intersections is
performed in Table \ref{Toverlap}.
\newcommand{\bra}{\text{[} }
\newcommand{\ket}{\text{]} }
\setlength{\extrarowheight}{3pt}
\begin{table}
\be
\begin{array}{|c|c|}
\hline \text{Field} & \text{Superpartner} \\
\hline
\tilde{\d}_m g_{ijk} & \bra  \pi_i, g_{ijk} \ket \\
\bra  g_{ijk},  \phi_{i} \ket    &  \bra g_{ijk},  \sigma_i  \ket
\\
 \bra g_{ijk},  E_i ] &  \bra  g_{ijk},   \rho_i  \ket  \\
 \bra \nu_{i} , g_{ijk} \ket  &  \bra g_{ijk},   \alpha_i \ket     \\
\hline
\end{array}
\ee
\mycaption{Degrees of freedom and their superpartners on ${\cal
U}_{ijk}$ from overlaps of fields defined on ${\cal U}_{ij}$. The
table has been obtained by operating $\partial_\lambda$ on Table
\ref{Doverlap}. \label{Toverlap}}
\end{table}
In all of these cases the \v{C}ech coboundary on ${\cal U}_{ijk}$
is merely the non-Abelian flux associated to an underlying field,
and would not seem to indicate the presence of additional
degrees of freedom.

\bigskip

To summarise, the local BRST operator of the twisted $N=4$ theory
on a local patch ${\cal U}_{i}$ does not involve {\em a priori}
any of the cocycle data $(g_{ijk},\lambda_{ij})$ in its
definition. If the underlying structure is not well-defined as a
principal  bundle but rather as a non-Abelian gerbe, we need to
consider the gauge theory on intersections of these local
descriptions separately. Then the automorphisms $\lambda_{ij}$
appear in the definition of twisted fields on the double
intersections ${\cal U}_{ij}$, and the group-element $g_{ijk}$
appears as a consequence of this as the ``discrepancy'' in the
three different twisted theories on ${\cal U}_{ijk}$.


\section{S-duality and self-duality}
\label{Duality}

The global structure of the non-Abelian gerbe is much looser than that of a principal bundle. This allows us to make use of some of the full quantum structure of the $N=4$ Yang-Mills theory in finding globally well-defined configurations.  The idea is that local descriptions on different charts ${\cal U}_{i}$ and ${\cal U}_{j}$ may be related by a non-perturbative symmetry of the theory, such as S-duality. This category of solutions of the quantum theory is related to non-geometric backgrounds \cf \cite{Hull:2004in}.

Apart from describing a new category of twisted $N=4$  Yang-Mills
configurations, this will contribute in developing intuition of
the physical significance of the fields that characterise a
non-Abelian gerbe, namely the curvature triple that consists of
\begin{itemize}
\item[{\bf --}] The {\em curvature} $\omega_{i} \in \Omega^{3}({\cal U}_{i}, \Lie G)$
\be
\omega_{i} = \d_{m_{i}} B_{i} ~;
\ee
\item[{\bf --}] The {\em intermediate curvature} $\delta_{ij} \in \Omega^{2}({\cal U}_{ij}, \Lie G)$
\be
\delta_{ij} = \lambda_{ij}(B_{j})-B_{i} + \d_{m_{i}} \gamma_{ij} - \half [\gamma_{ij}, \gamma_{ij}] ~;
\ee
\item[{\bf --}] The {\em fake curvature} $\nu_{i} \in \Omega^{2}({\cal U}_{i}, \Lie \Aut G)$
\be
\nu_{i} = \kappa(m_{i}) - \iota_{B_{i}} ~.
\ee
\end{itemize}
For properties of these differential forms, see \cite{Kalkkinen:2005py}.

The S-duality transformation $S$ acts on the complex coupling and
the field strength by
\be
\tau &\longrightarrow& -\frac{1}{\tau} \\
\kappa(m) &\longrightarrow& \star \kappa(m) ~.
\ee
As we do not concern ourselves with the action principle here, it
is only the latter that will be reflected in the structure of
the non-Abelian gerbe. Since local connections on different charts are related only
by the rather loose constraint
\be
{}^{\lambda_{ij}*}m_{j} - m_{i} + \iota_{\gamma_{ij}} &=& 0 ~,
\label{dgamma}
\ee
we can construct a non-geometric configuration where
\be
{}^{\lambda_{ij}} \kappa(m_{j}) &=& \star \kappa(m_{i}) \label{dr}
\ee
without implying too restrictive assumptions. (Here $*$ denotes
gauge transformation and $\star$ is the Hodge star.) When
$\lambda_{ij}$ is a trivial automorphism, this describes a
non-geometric background where the field $m_i$ on ${\cal
U}_i$ is the electric gauge potential, and the field $m_j$ on
${\cal U}_j$ is the magnetic gauge potential. They are directly
related to each other only at the intersection ${\cal U}_{ij}$,
where the difference is given by (\ref{dgamma}). The one-form $\gamma_{ij}$ appears as an effective gauge field on the double intersection when the intersection is interpreted as a defect. As $\lambda_{ij}$
acts on automorphisms by conjugation, traces remain invariant, and
the two instanton number densities on ${\cal U}_{ij}$ coincide.

\subsection{Consistency conditions}

To see what constraint (\ref{dr}) does imply, we should expand  it
as
\be
(1 - \star) \kappa(m_{i})|_{j} &=& \iota_{\d_{m_{i}}\gamma_{ij} - \half [\gamma_{ij}, \gamma_{ij}]} ~. \label{cons}
\ee
This means  that on the intersection ${\cal U}_{ij}$ we must be
able to write the fixed anti-self-dual two-form $(1 - \star)
\kappa(m_{i})|_{j}$ as an exact (combinatorial) differential of a
one-form $\gamma_{ij}$ as in the above formula (\ref{cons}). Note
that if $ \kappa(m_{i})$ happens to be purely self-dual, as is the
case for the solutions of the standard  twisted Yang-Mills theory,
we are at liberty to  choose the trivial solution $\gamma_{ij}=0$.

Imposing an analogue of (\ref{dr}) on every double  intersection
${\cal U}_{ij}$, ${\cal U}_{jk}$, and ${\cal U}_{ki}$ gives rise
to a consistency condition on their respective intersection ${\cal
U}_{ijk}$. There the situation depends on how $m_{k}$ is related
to $m_{i},m_{j}$. If we indeed assume  that the  duality relation
(\ref{dr}) holds  in every case $ij$, $jk$, and $ki$, we find using (\ref{parla}) that
the consistency condition (\ref{cons}) implies consistency on the
triple intersection as well
\be
\partial_{\lambda} \Big({}^{\lambda_{ij}} \kappa(m_{j}) - \star \kappa(m_{i})\Big) &=&  \partial_{\lambda} \tilde\delta_{m_{i}} \gamma_{ij} - \tilde\delta_{m_{i}}  \partial_{\lambda} \gamma_{ij} \\
&=& 0
\ee
in the notation of \cite{Kalkkinen:2005py}. Here
$\partial_\lambda$ is a $\lambda$-covariant \v{C}ech-differential (\ref{parla});
the check is that we can change charts $ij \longrightarrow jk
\longrightarrow ki$ in such a way that we come back to where we
started.

In this specific configuration on a Euclidean manifold ($\star^{2}
=1$) the commutator of the field strength with the class of the
gerbe $g_{ijk}$ is anti-self-dual
\be
(1 - \star) \kappa(m_{i})|_{jk} &=& [\kappa(m_{i}), \iota_{g_{ijk}} ] ~.
\ee
This means that, on triple intersections, the self-dual part of
every local field strength $\kappa(m_{i})|_{jk}$ commutes with
$g_{ijk}$. Hence, the class $g_{ijk}$ determines a local Abelian
system of self-dual fields in each  ${\cal U}_{ijk}$.  Suppose
next  that the two-form $B_{i}$ vanishes everywhere. Then the
curvatures of the non-Abelian gerbe summarise the construction
\be
\omega_{i} &=& 0 \\
\delta_{ij} &=& \d_{m_{i}}\gamma_{ij} - \half [\gamma_{ij}, \gamma_{ij}]  \\
\nu_{i} &=& \kappa(m_{i}) ~.
\ee
It is now not $\nu_{i}$ that is required to be self-dual as in twisted Yang-Mills, but rather $\delta_{ij}$. The present structure is therefore characterised by the following constraints
\be
B_{i} &=& 0 \\
\iota_{\delta_{ij}} &=& \star \iota_{\delta_{ij}} \\
{}^{\lambda_{ij}} \nu_{j} &=& \star \nu_{i} ~.
\ee

\subsection{The self-dual gerbe}

More generally, the above construction is  an example of self-dual
non-Abelian gerbes on four-manifolds  satisfying
\be
{\delta_{ij}} &=& \star {\delta_{ij}} \\
{}^{\lambda_{ij}} \nu_{j} &=& \star \nu_{i} ~.
\ee
The relation between the curvature triple  $(\omega_{i}, \delta_{ij},
\nu_{i})$ and the cocycle that classifies  the underlying gerbe
topologically $(g_{ijk}, \lambda_{ij})$ is as follows:
\begin{itemize}
\item[{\bf --}]  $\lambda_{ij}$ is the action of Hodge duality on fake curvature on ${\cal U}_{ij}$; and
\item[{\bf --}]  $g_{ijk}$ determines to what part of the Lie-algebra
$\ker \iota_{g_{ijk}}$ the self-dual part of $\nu_{i}$ is
restricted on ${\cal U}_{ijk}$.
\end{itemize}
These assumptions imply in particular $[\delta_{ij}, g_{ijk}]=0$.

The effect of allowing $B_{i}$ to be non-zero is to  relax the
anti-self-duality condition (\ref{dr}) somewhat, by subtracting an
inner automorphism part $\iota_{B}$ from the respective field
strengths that the condition relates. The curvature $\omega_{i}$
may now be non-zero, and measures precisely this departure from
the initial self-duality condition (\ref{dr}).

[As an aside, an other conceivable route of embedding this
non-geometric background in a gerbe would have been to set
$\delta_{ij} =0$, and parametrising the anti-self-dual part of
$\kappa(m_{i})$ by $B_{i}$
\be
(1 - \star) \kappa(m_{i})|_{j} &=& \iota_{B_{i} - \lambda_{ij}(B_{j}) } ~.
\ee
Under this assumption, however, consistency requires the vanishing of
\be
\partial_{\lambda} \Big({}^{\lambda_{ij}} \kappa(m_{j}) - \star \kappa(m_{i})\Big) = [g_{ijk}, \nu_{i}]
\ee
on triple intersections. This means that $g_{ijk}$ determines an Abelian frame for the restrictions of the {\em whole} fake curvature; such assumptions have the tendency of making the gerbe effectively  Abelian.]


\section{The flux of a non-Abelian gerbe}
\label{Flux}

In this section we shall first discuss how 't Hooft's  magnetic
flux appears  traditionally in Yang-Mills theory. This flux is
classified in $H^{2}(X,\Z G)$, and can be thought of in terms of
the change of  the gauge group from electric to magnetic. A
similar loosening of structure leads to magnetic flux associated
to the class of a non-Abelian gerbe in $H^{1}(X,  G \ltimes \Aut G
)$.

\subsection{'t Hooft's Abelian magnetic fluxes}
\label{Ab}

In  $N=4$ super-Yang-Mills all fields are in the adjoint
representation, and the gauge group is $G/\Z G$ rather than the
full exponential group of the Lie-algebra. We will consider in what follows the special unitary case of $G=\SU(n)/ \Zset_{n}$.  The magnetic dual of this group is the
full special unitary group $G^{\mathrm{v}}=\SU(n)$ with the centre
restored \cite{Godd}. (Another interesting example is the pair
$G=\Spin(8)$, $G^{\mathrm{v}}=\Spin(8)/\Zset_{2} \times
\Zset_{2}$. Same observations apply.)

Consider an ``electric'' principal bundle $E$ with transition
functions  $\transf_{ij}$ valued in the gauge group $G=\SU(n)/
\Zset_{n}$. Choose a lift from  $G=\SU(n)/ \Zset_{n}$ to
$G^{\mathrm{v}}=\SU(n)$. On a triple intersection ${\cal U}_{ijk}$
the lifted transition functions $\hat\transf_{ij}$ do not
necessarily satisfy the usual cocycle condition, but there may be
an obstruction
\be
\hat\transf_{ij} \hat\transf_{jk} \hat\transf_{ki} &=& a_{ijk}  ~,
\ee
where $a_{ijk} \in \Z G$. If $G=\SU(n)$, we can think of these Abelian obstructions in terms of $n\times n$ matrixes
\be
a_{ijk} &=& \e^{2\pi i \frac{k_{ijk}}{n}} \unit_{n} ~, \qquad
k_{ijk} \in \Zset~.
\ee

We may attempt to remove this obstruction by changing our choice
of lift  consistently on each  intersection
\be
\hat\transf_{ij}' &=& \hat\transf_{ij} k_{ij} ~, \qquad k_{ij} \in \Z G   ~.
\ee
If it turns out that the mismatch  $a_{ijk}$ cannot be compensated
for by changing the lift in this way, we have a true obstruction
$[a_{ijk}] \in \check{H}^{2}(X,\Zset_{n})$ to the lift. On the
other hand, if it turns out that $[a_{ijk}]=0$, then the lifted
bundle $\hat{E}$ exists as a globally well-defined entity.

One may look for such obstructions
\cite{'tHooft:1977hy,'tHooft:1979uj}  by calculating Wilson loops
along closed paths. The magnetic flux captured inside the loop is
precisely the above obstruction $[k_{ijk}]$. (The exponential of
this $[a_{ijk}]$ is rather the surface holonomy associated to this
magnetic flux; they both describe the same physics.) In the
electric picture we have therefore a well-defined $G$-bundle $E$.
In the magnetic picture no such global $G^{\mathrm{v}}$-bundle
exists unless  the (torsion class) magnetic flux $[a_{ijk}]$
vanishes. If it does not vanish, the global structure on the
magnetic side is a flat Abelian gerbe, rather than a principal
$G^{\mathrm{v}}$-bundle. This magnetic flux  satisfies the cocycle
condition
\be
a_{jkl} a_{ijl} = a_{ijk}a_{ikl}  ~.  \label{co1}
\ee

\subsection{Outer automorphisms}

Suppose we are given locally a well-defined principal $G$-bundle
$P_{i}$  on each local neighbourhood ${\cal U}_{i}$, and
invertible mappings $\lambda_{ij}: P_{j} \longrightarrow P_{i}$
that act by automorphisms $\Aut G$ on the fibre $G$. The
automorphisms need not be just conjugations by a group element,
but could well be outer automorphisms, such as complex conjugation
for $G=\SU(n)$, or triality for $G=\Spin(8)$.

Given a general automorphism $\lambda_{ij}$, there is no universal
split to inner and outer automorphisms. As the latter are defined as the quotient $\Aut G/ \Int G = \Out G$,  we can nevertheless
project an automorphism to its outer part $p(\lambda_{ij}) =
w_{ij}$. Suppose we are given such a  pure outer automorphism on each intersection ${\cal U}_{ij}$ that satisfies
\be
w_{ij}w_{jk}w_{ki} &=& \unit ~. \label{cons2}
\ee
This amounts to choosing a class
\be
[w] \in H^{1}(X, \Out G) ~,
\ee
and determines a principal $\Out G$-bundle in $\Tors \Out G$. The most obvious example of this structure is perhaps Yang-Mills on a local $\Out  G$-orbifold. Then, the surface holonomy of an Abelian gerbe picks up discrete torsion that can be understood in precisely these terms \cite{Sharpe:2000ki}.

If we lift these outer automorphisms from $\Out G$ to $\lambda_{ij} \in \Aut G$, the consistency condition (\ref{cons2}) is replaced by
\be
 \lambda_{ij}\lambda_{jk} \lambda_{ki} &=& \iota_{g_{ijk}}
\ee
for some mapping to the group $g_{ijk} : {\cal U}_{ijk} \longrightarrow G$. This will complicate the cocycle condition satisfied by $g_{ijk}$, however.

Consider the case $G=\SU(n)$, so that $\Z G$ are given by $n$th roots
of unity. Suppose that $p(\lambda_{ij})$ acts by complex conjugation, and $\lambda_{jk}$, $\lambda_{ki}$ are pure conjugations. Because the complex conjugation will act also on $\lambda_{jk}$, $\lambda_{ki}$ in the definition of $\iota_{g_{ijk}}$,  it is clear that a direct analogue of the Abelian cocycle condition (\ref{co1}) will not be satisfied. The appropriate generalisation will lead us to the topic of the next section:

\subsection{Crossed modules}
\label{CM}

The problem of generalising (\ref{co1}) to an equation that could be valid also for non-Abelian cocycles can be solved, when  a way to keep track of the ``frame'' in which a group element is given is developed. The right structure for this is the {\em crossed module}: This structure consists of the groups $G$ and $H$,  the homomorphism $\partial : G \longrightarrow H$ and the action of $h \in H$ on $g,g' \in G$ denoted \eg by $g \mapsto {}^{h}g$. The homomorphism $\partial$ is required to satisfy
\be
\partial ({}^{h}g) &=& \iota_{h}(\partial g) \\
{}^{\partial g}(g') &=& \iota_{g} (g') ~.
\ee
We shall be interested in the case $H = \Aut G$ when the homomorphism $\partial = \iota$ is the conjugation by a group element.

We will consider, in particular, the group-valued function
$g_{ijk} \in G$ on ${\cal U}_{ijk}$ and the automorphism-valued
function $\lambda_{ij} \in \Aut G$ on ${\cal U}_{ij}$. This pair
$(g_{ijk},\lambda_{ij})$ defines locally an element of the crossed
module $G \ltimes \Aut G$.  The cocycle equations \cite{BreenAst}
that they satisfy are
\be
\lambda_{ij}(g_{jkl}) g_{ijl} &=&g_{ijk}g_{ikl}   \\
\iota_{g_{ijk}}\lambda_{ik} &=& \lambda_{ij}\lambda_{jk} ~.
\ee
Two equivalent cocycles $( g_{ijk},\lambda_{ij})$ and $(
g_{ijk}',\lambda_{ij}')$ differ by a coboundary; the coboundary
equations \cite{BreenAst,BM2} are quite involved due to the fact
that for writing them down one should decompose the gerbe fully.
In fact, the data that goes in this decomposition is effectively
the data that is included in the differential geometry of such a
fully-decomposed gerbe.

When these equivalencies are taken in account correctly, such a cocycle pair (modulo the coboundary relations) determines a cohomology class of a non-Abelian gerbe
\be
[(g_{ijk},\lambda_{ij})] &\in& H^{1}(X, G \ltimes \Aut G) ~.
\ee
This group is the direct generalisation of the \v{C}ech-cohomology group $H^{1}(X,G)$ whose elements determine isomorphism classes of principal $G$-bundles, \ie $\Tors G$. We shall denote $\mathbf{G} = G \ltimes \Aut G$. Sometimes also the notation $G \stackrel{\iota}{\longrightarrow} \Aut G$ is used as it emphasises the r\^ole played by the homomorphism $\iota$.

We have already  encountered two examples of such a cocycle,
namely the Abelian magnetic flux $(a_{ijk},\unit)$, and the outer
automorphisms $(\unit,w_{ij})$. More generally, the cohomology
group $H^{1}(X, \mathbf{G})$ of a non-Abelian gerbe fits in the
exact sequence \cite{Breen}
\be
H^{0}(X, \Out G)  \longrightarrow H^{2}(X, \Z G) \longrightarrow
H^{1}(X, \mathbf{G}) \longrightarrow \Tors(\Out  G) ~. \label{es}
\ee
The image of elements $[g_{ijk}] \in H^{2}(X, \Z G) $ in $H^{1}(X,
\mathbf{G})$ is $(g_{ijk},\unit)$;  the image of a general
$(g_{ijk},\lambda_{ij})$ in $\Tors(\Out  G)$ is in the equivalence
class of principal bundles given by $[p(\lambda)] \in H^{1}(X, \Out G)$.

A category of examples that carry this non-Abelian generalisation
$[(g_{ijk},\lambda_{ij})]$ of the more usual Abelian magnetic flux
$[a_{ijk}]$ would be orbifold theories where the orbifold action $
\lambda_{ij}$ involves an arbitrary conjugation with a group
element, and not just the outer part of the automorphism group.
These theories are locally $N=4$ supersymmetric outside the actual
fixed point locus.


\section{Chiral five-branes}
\label{Chir}

I will include in this section a few remarks on eventual applications
of the above observation on describing partially the worldvolume
dynamics of a stack of chiral five-branes.

The uncompactified six-dimensional worldvolume theory for a single
chiral M-theory five-brane involves the $N=(0,2)$ tensor
multiplet \cite{Howe:1983fr}. The tensor field couples
to tensionless worldvolume strings whose dynamics give the
parallel low-energy excitations of the worldvolume; the five
scalar fields in the multiplet give the transverse excitations. At
weak worldsheet coupling a stack of these branes has the
worldvolume excitations of the Little String Theory
\cite{Aharony:1998ub}.

Geometrically the two-form can be thought of as a connection on an
Abelian gerbe on the worldvolume \cite{Dijkgraaf:1998xr}. The
Deligne class of the gerbe on the brane is twisted by the class of
the bulk two-gerbe \cite{Kalkkinen:2004hs} in a direct analogue to
what happens in String Theory \cite{Freed:1999vc}. This is also
how the elusive $E_{8}$ structure \cite{Diaconescu:2000wy} enters
the geometry of gauge fields in M-theory \cite{Aschieri:2004yz}.
It seems therefore reasonable that the low-energy dynamics of a
stack of these branes should be described geometrically by a
non-Abelian gerbe. The matter turns out to be much more subtle
than that, owing \eg to the inherent non-localities on the
non-critical worldvolume string theory.

Reduced from six to four dimension, the tensor multiplet reduces
however  to the $N=4$ vector multiplet. A reduction of an M-theory
five-brane on a torus, in particular, can be related directly to
the self-dual D3-brane \cite{Berman:1998va} whose low-energy
description is the $N=4$ supersymmetric Yang-Mills theory.
Wrapping the brane around a more general holomorphic cycle
$\Sigma$ breaks supersymmetry by a further half, and one obtains
the four-dimensional $N=2$ super-Yang-Mills theory. The
four-dimensional interpretation of the cycle $\Sigma$ is that it
is the Seiberg-Witten curve \cite{Witten:1997sc}.

Consider a five-brane $M$ that  is locally of the form ${\cal
U}_{i} \times \Tset^{2}$, where $\{ {\cal U}_{i} \}$ is a cover of
a Euclidean four-manifold $X$. On each ${\cal U}_{ij}$  the
$\Tset^{2}$-fibres can be related one to an other by
$\SL(2,\Zset)$ transformations that act precisely as S-duality
transformations on the remaining degrees of freedom on $X$. As the
worldvolume degrees of freedom are tensionless strings, there are
massless winding modes in any limit we might consider. In the
large $\Vol \Tset^{2}$ limit Kaluza-Klein modes are suppressed,
however, and we get an (approximative) transverse $\SO(6)$
invariance in eleven dimensions.

The transverse $\SO(6)$ symmetry together with the fact that  the
five-brane breaks half of the supersymmetries in the bulk mean
that the effective theory on $X$ includes the $N=4$
super-Yang-Mills theory. Even if this local quantum field theory
misses some of the remaining massless non-local degrees of freedom
on the five-brane, it is nevertheless a unitary quantum field
theory, and we can consider it as a self-consistent sub-sector of
the full worldvolume theory on $M$.

In a flat or hyper-K\"ahler background metric the twisted and  the
physical $N=4$ super-Yang-Mills theory are equivalent
\cite{Witten:1994ev,Vafa:1994tf}. The observations in this paper
can therefore be applied to five-branes on six-manifolds that are
torus bundles over some hyper-K\"ahler manifold, such as K3 or an
ALE space in the non-compact case. Abelian gerbes on toric
fibrations and string compactifications on stacks have been
discussed in the Abelian case in
\cite{Donagi:2003av,Pantev:2005wj}. For a discussion on Conformal
Field Theory and branes, see \cite{Gawedzki:2004tu}.

\bigskip

A more direct relationship between the  tensionless tensor theory
and Yang-Mills could arise already in five dimensions;
these twisted theories have not been worked out in detail,
however. Indeed, a reduction of the five-brane theory on a circle
yields the five-dimensional super-Yang-Mills theory with 16
supercharges. The spin-groups relevant to this theory are
\be
\begin{array}{rcclcrccl}
\Spin^{0}(4,1) &=& \Sp_{1,1} & ~ \subset ~ & \Spin^{0}(5,1) &=& \SL(2, \Hset) \\
\Spin(5) &=& \Sp_{2} & ~ \subset  ~ & \Spin(6) &=& \SU(4)
\end{array} ~.
\ee
In the Euclidean case the worldvolume spin group and the
R-symmetry  group are both $\Sp_{2}$, and we can twist the theory
by identifying the two. This will give rise to a Fermionic
two-form, a vector, and a scalar (\eg ``$\alpha,\pi,\sigma$'')
from gaugini, and a Bosonic vector, say  $b$, from scalars. A
vector (in the Abelian case) is dual to a two-form in five
dimensions $\d b =
* \d B$.

It is interesting to note that if we had at our disposal a
determinant-like  homomorphism ${\det}_{\Hset} : \Sp_{1,1}
\longrightarrow \SU(2)$ \cite{Quat}, we could use it to twist the
five-dimensional theory with the diagonal subgroup $\SU(2) \subset
\Sp_{2}$ of the R-symmetry group such that the R-symmetry
representation of the four supercharges splits $\mathbf{4}
\longrightarrow \mathbf{2} \oplus \mathbf{2}$. As the scalars are
in the antisymmetric $\mathbf{5}$ of $\Sp_{2}$ this means that
they decompose to $\mathbf{3} \oplus \mathbf{1} \oplus
\mathbf{1}$. The four-dimensional interpretation of this matter
content is a self-dual two-form and two scalars --- the third
scalar needed in the four-dimensional theory arises from the
reduction of the gauge field. However, from the five-dimensional
point of view this decomposition is also that of a massive vector
field: it might be interesting to look for a massive version of
the above Hodge duality, and a relationship to a (massive) tensor
field in five dimensions.


\section{Discussion}
\label{Disc}

The link between the BRST operator of a non-Abelian gerbe and
twisted Yang-Mills theory allows a generalisation of Yang-Mills
theory where local structures are related to each other in a
looser fashion than in a standard principal bundle. Where the
local structure of standard Yang-Mills theory is determined by a
gauge equivalence class of the transition functions $h_{ij}$, the
data needed in this generalised structure is an element of a
crossed module $(g_{ijk},\lambda_{ij})$ determining a class in
$H^{1}(X, \mathbf{G})$.

As the class of the gerbe depends both on $g_{ijk}$ and
$\lambda_{ij}$, these quantities do not really have invariant
meaning separately. If we have chosen a specific representative
$(g_{ijk},\lambda_{ij})$ of a class $[(g_{ijk},\lambda_{ij})] \in
H^{1}(X, \mathbf{G})$, we may nevertheless try to see what the
physical origin of these two quantities is. As explained in the
Paper, $\lambda_{ij}$ can be thought of as a generalisation of the
transition functions in Yang-Mills theory, and $g_{ijk}$ can be
thought of as a non-Abelian generalisation of magnetic flux. Such
an Abelian magnetic flux $a_{ijk}$ showed up in lifting the
transition functions to the magnetic gauge group
\be
\widehat{s_i {s}_j^{-1}} = \hat{h}_{ij}
\ee
and comparing them over a triple intersection
\be
\partial (\widehat{s_i {s}_j^{-1}}) &=& a_{ijk} ~;
\ee
similarly, if we have three independent differential forms
$\phi_i$, $\phi_j$, and $\phi_k$ defined over the same point in
${\cal U}_{ijk}$ in a non-Abelian gerbe, the respective
discontinuities on ${\cal U}_{ij}$, ${\cal U}_{jk}$, and ${\cal
U}_{ki}$ satisfy
\be
\partial_\lambda \Big( \lambda_{ij}(\phi_j) \phi_i^{-1} \Big) &=& [g_{ijk},
\phi_i]~.
\ee
(We use the multiplicative notation of combinatorial differential
geometry to emphasise the analogy.)

Magnetic flux in a standard Yang-Mills theory leads of course to a
milder loosening of the electric structure of the theory. This
flux can be classified in terms of centre valued classes in
$H^{2}(X, \Z G)$. The invariant statement
is that the class of the gerbe $[(g_{ijk},\lambda_{ij})]$
generalises that Abelian magnetic flux $[a_{ijk}]$ to a
non-Abelian context in the sense of the exact sequence (\ref{es}).

\bigskip

The local structure of the thus loosened theory gives rise to new
degrees of freedom localised on double intersections of local
charts, where two conflicting theories overlap. We have argued
that it is useful to think of these overlaps as domainwalls with
dynamics given by fields either switched off in the bulk theory
($C^{\Tw}$ and $\zeta^{\Tw}-\eta^{\Tw}$) or arising from the
mismatch of the local fields in the two neighbourhoods
($\gamma_{ij}$). Technically this required choosing a
triangulation of $X$ compatible with the cover we use $\{ {\cal
U}_i \} $, and attaching an index $i$ to each volume in the
triangulation, $ij$ codimension one simplex, $ijk$ codimension two simplex, and so on. Then the domainwall degrees of freedom are indeed localised in a (network) of simplexes labelled by index pairs $ij$.

It was further argued that there are no new degrees of freedom in
the codimension two simplexes $\Sigma_{ijk} \subset {\cal U}_{ijk}$ labelled by index triples $ijk$. These are generically codimension two surfaces, and correspond in four dimensions to (Euclidean) string worldsheets. The class of the gerbe involves nevertheless the fixed  mappings $g_{ijk}: {\Sigma}_{ijk} \longrightarrow G$. 
As argued above, this map is a generalisation of the Abelian
magnetic flux that arises as the centre part of a Wilson
line in magnetic configurations. It plays therefore naturally 
the r\^ole of a surface holonomy of the non-Abelian gerbe over the surface $\Sigma_{ijk}$. This generalisation requires, of course,  revising what usually is meant by a surface holonomy, and somewhat side-steps problems arising in more direct definitions of surface holonomies that depend on a choice of surface ordering \eg
\cite{Akhmedov:2005tn,Gustavsson:2005fp}. For generalisations that
make use of non-Abelian two-forms see \eg
\cite{Hofman:2002ey,Baez:2004in}.

These domainwalls can be  identified in fact with membranes
moving inside the four-dimensional bulk space. The worldvolume
theory on them is indeed always the super-Yang-Mills theory
reduced from ten dimensions, in this case {\em via} the twisted
four-dimensional theory.  Due to the topological nature of these
membranes, one might suspect that they are related to the
topological Dirac branes $U_{3}$ on the five-brane worldvolume
whose boundaries are the tensionless worldvolume strings $W_{2} =
\partial U_{3}$ in the notation of Ref.~\cite{Kalkkinen:2002tk}.
This structure is in fact required in order to embed a stack of
interacting membranes in the five-brane worldvolume.

\bigskip

Apart from the non-Abelian fluxes and the defect  dynamics, an
other new aspect in quantum field theory is how the non-local
structure of the fields on the gerbe generalises the global
structure of the twisted theory: Indeed, on a triple intersection
we were consequently forced to consider three different scalar
fields $C^{\Tw}$, whose covariant difference --- in the sense
explained in (\ref{parla}) --- was related to one of the local
fields on the gerbe $\phi_{i}$ and the cocycle data $g_{ijk}$.
This non-local structure came into its own when considering
non-geometric backgrounds where gauge fields on adjacent charts
were related by S-duality. It was possible to give an explicit
formula for this relation  consistently off-shell, as the
relationship between the gauge fields was determined up to an
arbitrary  group-valued one-form $\gamma_{ij}$. It turned out that
the r\^ole of the cocycle $g_{ijk}$ in this case was to constrain
the self-dual part on triple intersections.

As the ghost number assignments in the gerbe  and in the twisted
theory are different, the action principle will not be the same.
In want of an action principle we have not been in a
position to check that the above-mentioned non-geometric
background reduces to the expected electric-magnetic dual
background also on-shell. This matter should clearly be clarified,
as well as the construction of actions in general.

An other consequence of the difference in ghost number assignment is the
fact that the topological Yang-Mills theory on the gerbe is not
balanced, and that the partition function is therefore not
protected from ghost number anomalies. This is interesting in view
of  constructing observables \cite{Kalkkinen:2005py}.

These observations have immediate implications for  the study of
the geometry of chiral five-branes. Though the $N=4$
supersymmetric Yang-Mills theory captures only a part of the
dynamics in those systems, the present generalisation allows the
inclusion of some of the expected non-local phenomena in the field
theory discussion, such as those related to Dirac membranes ending
on tensionless strings, and the holonomies associated to the
worldsheets of these strings.

\subsubsection*{Acknowledgements}
I would like to thank Larry Breen, Chris Hull, and Bernard Julia
for discussions,  and Urs Schreiber and Eric Sharpe for
correspondence.   This research is supported by a Particle Physics
and Astronomy Research Council (PPARC) Postdoctoral Fellowship.


\begin{thebibliography}{99}

\bibitem{Kalkkinen:1999uz}
  J.~Kalkkinen,
  ``Gerbes and Massive Type II Configurations,''
  JHEP {\bf 9907} (1999) 002
  [{\tt hep-th/9905018}].

\bibitem{Kalkkinen:2005py}
  J.~Kalkkinen,
  ``Topological Quantum Field Theory on Non-Abelian Gerbes,''
  [{\tt hep-th/0510069}].

\bibitem{Yamron:1988qc}
  J.P.~Yamron,
  ``Topological Actions from Twisted Supersymmetric Theories,''
  Phys.\ Lett.\ B {\bf 213} (1988) 325.

\bibitem{Vafa:1994tf}
  C.~Vafa and E.~Witten,
  ``A Strong Coupling Test of S-Duality,''
  Nucl.\ Phys.\ B {\bf 431} (1994) 3
  [{\tt hep-th/9408074}].

\bibitem{Labastida:1997vq}
  J.M.F.~Labastida and C.~Lozano,
  ``Mathai-Quillen Formulation of Twisted $N = 4$ Supersymmetric
  Gauge Theories in Four Dimensions,''
  Nucl.\ Phys.\ B {\bf 502} (1997) 741
  [{\tt hep-th/9702106}].

\bibitem{Hull:2004in}
  C.M.~Hull,
  ``A Geometry for Non-Geometric String Backgrounds,''
  [{\tt hep-th/0406102}].

\bibitem{Freed:1999vc}
D.S.~Freed and E.~Witten, ``Anomalies in String Theory with
D-Branes,'' [{\tt hep-th/9907189}].

\bibitem{Kalkkinen:2002tk}
J.~Kalkkinen and K.~S.~Stelle, ``Large Gauge Transformations in
M-theory,'' J.\ Geom.\ Phys.\  {\bf 48}, 100 (2003)   [{\tt
hep-th/0212081}].

\bibitem{Kalkkinen:2004hs}
J.~Kalkkinen, ``Holonomies of Intersecting Branes,'' Fortsch.\
Phys.\  {\bf 53} (2005) 913 [{\tt hep-th/0412166}].

\bibitem{'tHooft:1977hy}
  G.~'t Hooft,
  ``On the Phase Transition towards Permanent Quark Confinement,''
  Nucl.\ Phys.\ B {\bf 138} (1978) 1.

\bibitem{'tHooft:1979uj}
  G.~'t Hooft,
  ``A Property of Electric and Magnetic Flux in Non-Abelian
  Gauge Theories,''
  Nucl.\ Phys.\ B {\bf 153} (1979) 141.

\bibitem{Godd}
  P.~Goddard, J.~Nuyts and D.I.~Olive,
  ``Gauge Theories and Magnetic Charge,''
  Nucl.\ Phys.\ B {\bf 125} (1977) 1.

\bibitem{Kapustin:2005py}
  A.~Kapustin,
  ``Wilson-'t Hooft Operators in Four-Dimensional Gauge Theories and
  S-Duality,''
  [{\tt hep-th/0501015}].

\bibitem{Witten:1988ze}
  E.~Witten, ``Topological Quantum Field Theory,'' Commun.\ Math.\
  Phys.\  {\bf 117} (1988) 353.

\bibitem{Marcus:1995mq}
  N.~Marcus,
  ``The Other Topological Twisting of $N=4$ Yang-Mills,''
  Nucl.\ Phys.\ B {\bf 452} (1995) 331
  [{\tt hep-th/9506002}].

\bibitem{Witten:1994ev}
E.~Witten, ``Supersymmetric Yang-Mills Theory on a
Four-Manifold,'' J.\ Math.\ Phys.\  {\bf 35} (1994) 5101 [{\tt
hep-th/9403195}].

\bibitem{BreenAst}
  L.~Breen,  ``Classification of 2-Gerbes and 2-Stacks,''
  Ast\'erisque {\bf 225}, Soci\'et\'e Math\'ematique de France (1994).

\bibitem{BM2}
  L.~Breen and W.~Messing, ``Differential Geometry of Gerbes,''
  [{\tt math.ag/0106083}].

\bibitem{M} M.K.~Murray, ``Bundle Gerbes,''
  J.~Lond.~Math.~Soc.~{\bf 54} (1996) 403.

\bibitem{BM1}
  L.~Breen and W.~Messing, ``Combinatorial Differential Forms,''
  [{\tt math.ag/0005087}].

\bibitem{Dijkgraaf:1996tz}
  R.~Dijkgraaf and G.~W.~Moore,
  ``Balanced Topological Field Theories,''
  Commun.\ Math.\ Phys.\  {\bf 185} (1997) 411
  [{\tt hep-th/9608169}].

\bibitem{Brylinski} J.L.~Brylinski, ``Loop Spaces,
Characteristic Classes and Geometric Quantization,''(Birk\"auser,
Boston 1993)

\bibitem{Sharpe:2000ki}
  E.R.~Sharpe,
  ``Discrete Torsion,''
  Phys.\ Rev.\ D {\bf 68} (2003) 126003
  [{\tt hep-th/0008154}].

\bibitem{Breen}
  L.~Breen, ``Bitorseurs et Cohomologie Non Ab\'elienne,'' in The
  Grothendieck Festschrift, Progress  in Mathematics {\bf 86},
  Birkh\"auser, 401-476 (1990).

\bibitem{Howe:1983fr}
  P.S.~Howe, G.~Sierra and P.K.~Townsend,
  ``Supersymmetry in Six Dimensions,''
  Nucl.\ Phys.\ B {\bf 221} (1983) 331.

\bibitem{Aharony:1998ub}
  O.~Aharony, M.~Berkooz, D.~Kutasov and N.~Seiberg,
  ``Linear Dilatons, NS5-Branes and Holography,''
  JHEP {\bf 9810} (1998) 004
  [{\tt hep-th/9808149}].

\bibitem{Dijkgraaf:1998xr}
  R.~Dijkgraaf,
  ``The Mathematics of Five-Branes,''
  [{\tt hep-th/9810157}].

\bibitem{Diaconescu:2000wy}
D.E.~Diaconescu, G.W.~Moore and E.~Witten, ``$E_{8}$ Gauge Theory,
and a Derivation of K-Theory from M-Theory,'' Adv.\ Theor.\ Math.\
Phys.\  {\bf 6}, 1031 (2003) [{\tt hep-th/0005090}].

\bibitem{Aschieri:2004yz}
P.~Aschieri and B.~Jurco, ``Gerbes, M5-Brane Anomalies and $E_{8}$
Gauge Theory,'' JHEP {\bf 0410}, 068 (2004) [{\tt
hep-th/0409200}].

\bibitem{Berman:1998va}
D.~Berman, ``M5 on a Torus and the Three-Brane,'' Nucl.\ Phys.\ B
{\bf 533}, 317 (1998) [{\tt hep-th/9804115}].

\bibitem{Witten:1997sc}
E.~Witten, ``Solutions of Four-Dimensional Field Theories via
M-Theory,'' Nucl.\ Phys.\ B {\bf 500} (1997) 3 [{\tt
hep-th/9703166}].

\bibitem{Donagi:2003av}
  R.~Donagi and T.~Pantev,
  ``Torus Fibrations, Gerbes, and Duality,''
  [{\tt math.ag/0306213}].

\bibitem{Pantev:2005wj}
  T.~Pantev and E.~Sharpe,
  ``String Compactifications on Calabi-Yau Stacks,''
  [{\tt hep-th/0502044}].

\bibitem{Gawedzki:2004tu}
  K.~Gawedzki,
  ``Abelian and Non-Abelian Branes in WZW Models and Gerbes,''
  Commun.\ Math.\ Phys.\  {\bf 258} (2005) 23
  [{\tt hep-th/0406072}].

\bibitem{Quat}
N.~Cohen and S.~de Leo, ``The Quaternionic Determinant,''
Electr.~J.~Lin.~Alg.~{\bf 7} (2000) 100.

\bibitem{Akhmedov:2005tn}
  E.T.~Akhmedov,
  ``Towards the theory of non-Abelian tensor fields. I,''
  [{\tt hep-th/0503234}].

\bibitem{Gustavsson:2005fp}
  A.~Gustavsson,
  ``A reparametrization invariant surface ordering,''
  [{\tt  hep-th/0508243}].

\bibitem{Hofman:2002ey}
  C.~Hofman,
  ``Nonabelian 2-Forms,''
  [{\tt hep-th/0207017}].

\bibitem{Baez:2004in}
  J.~Baez and U.~Schreiber,
  ``Higher Gauge Theory: 2-Connections on 2-Bundles,''
  [{\tt hep-th/0412325}].

\end{thebibliography}
\end{document}